\documentclass[aps,prb,a4paper,10pt,twocolumn,amsmath,showpacs,superscriptaddress]{revtex4-2} 
\usepackage[hmargin=2cm,vmargin=2.5cm]{geometry}

\usepackage[colorlinks,plainpages=false,linkcolor=blue,urlcolor=blue,citecolor=blue,pdfpagemode=UseNone,pdfstartview=FitBH]{hyperref}
\usepackage[charter,greekuppercase=italicized]{mathdesign}
\usepackage{xcolor} 
\usepackage{graphicx}    
\graphicspath{{./}{figure/}}
\DeclareGraphicsExtensions{.eps,.png,.pdf,.jpg}

\newcommand{\casn}{Ca\-Sn$_{3}$}

\def\ca{CaSn$_{3}$}
\def\ac3{AuCu$_{3}$}
\def\tc{$T_{c}$}

\def\hc2{$H_{c2}$}

\def\cbi{Cu$_{x}$Bi$_{2}$Se$_{3}$}

\def\mbi{$M_{x}$Bi$_{2}$Se$_{3}$}
\def\oh{$O_{h}$}
\def\para{\parallel}
\def\musr{$\mu$SR}
\def\zfmu{ZF-$\mu$SR}
\def\tfmu{TF-$\mu$SR}

\definecolor{ao}{rgb}{0.0, 0.5, 0.0}

\begin{document}

\title{Nematic superconductivity in the topological semimetal CaSn$_{3}$}							 

\author{K. A. M. H. ~Siddiquee}
\author{R.~Munir}
\author{C.~Dissanayake}
\author{P.~Vaidya}
\author{C.~Nickle}
\author{E.~Del Barco}
\affiliation{Department of Physics, University of Central Florida, Orlando, Florida 32816}
\author{G.~Lamura}
\affiliation{CNR-SPIN, Corso Perrone 24, 16152 Genova, Italy}
\affiliation{Dipartimento di Fisica, Universit\`{a} di Genova, 16146 Genova, Italy}
\author{C.~Baines}
\affiliation{Laboratory for Muon-Spin Spectroscopy, Paul Scherrer Institut, CH-5232 Villigen PSI, Switzerland}
\author{S. Cahen}
\author{C.~H\'{e}rold}
\affiliation{Institut Jean Lamour, UMR 7198 CNRS-UL, 2 All\'{e}e Andr\'{e} Guinier, B.P. 50840, 54011 Nancy Cedex, France}
\author{P.~Gentile}
\affiliation{CNR-SPIN, I-84084 Fisciano (Salerno), Italy}
\affiliation{Dipartimento di Fisica "E.\ R.\ Caianiello", Universit\`a di Salerno, I-84084 Fisciano (Salerno), Italy}
\author{T.~Shiroka}
\affiliation{Laboratory for Muon-Spin Spectroscopy, Paul Scherrer Institut, CH-5232 Villigen PSI, Switzerland}
\affiliation{Laboratorium f\"ur Festk\"orperphysik, ETH-H\"onggerberg, CH-8093 Z\"urich, Switzerland}
\author{Y.~Nakajima}
\email[Corresponding author: ]{Yasuyuki.Nakajima@ucf.edu}
\affiliation{Department of Physics, University of Central Florida, Orlando, Florida 32816}


\date{\today}

\begin{abstract}
The superconducting behavior of the topological semimetal {\ca} was 
investigated by means of magnetotransport and muon spectroscopy (\musr) 
measurements, both providing strong evidence of \emph{nematic} behavior. 
Magnetotransport detects an anisotropic upper critical field,  
characterized by a twofold symmetry about $C_{4}$ axis, thus breaking 
the rotational symmetry of the underlying cubic lattice.
Transverse-field {\musr} data support such picture, with the 
muon depolarization rate depending strongly on the magnetic field direction,  
here applied along the [110] or [001] crystal directions. 
In the former case, the absence of any additional muon depolarization suggests an \emph{unconventional} vortex lattice. In the latter case, a vortex lattice encompassing a sample volume of at least 52\% indicates the bulk nature of CaSn$_{3}$ superconductivity. 
The resulting superfluid density in the (001) planes shows a gapped 
low-temperature behavior, with a superconducting gap value $\Delta(0)\simeq 0.61(7)$\,meV. Additional zero-field \musr\ results indicate that the superconducting state 
is time-reversal-invariant. 
This fact and the breaking of rotational symmetry in a fully-gapped superconductor 
are consistent with an unconventional pairing state in a multi-dimensional 
representation, thus making {\ca} an important example of nematic superconductor.
\end{abstract}


\maketitle

\section{Introduction}

Topological superconductors (TSCs), hosting Majorana fermions on their boundary, 
have  
attracted considerable attention due to their potential use in quantum computing \cite{schny08}. Topological superconducting states are closely linked with pairing symmetry and Fer\-mi\--sur\-face topology. In time-reversal-invariant and inversion-symmetric systems, topological superconductivity requires odd-parity pairing with a fully-opened superconducting gap,  described by $\Delta(-\boldsymbol{k})=-\Delta(\boldsymbol{k})$, and Fermi surfaces that enclose an odd number of time-reversal-invariant momenta (TRIM) \cite{fu10,fu14,vende16a}. In spin-rotational-invariant systems, odd-parity superconductivity arises from spin-triplet pairing, mediated by ferromagnetic instabilities, as observed in UPt$_{3}$ \cite{joynt02}. On the other hand, in spin-orbit-coupled systems, odd-parity pairing can be stabilized even in the absence of magnetic instabilities \cite{fu10}.

Thanks to the presence of a strong spin-orbit coupling (SOC), the doped Bi-based topological insulator system {\mbi} ($M$ = Cu, Sr, Nb) is among the most extensively studied TSC candidate materials \cite{hor10a,krien11,shrut15}. In these compounds, the topological nature of the electronic structure does not depend on doping, as confirmed by angle resolved photoemission spectroscopy (ARPES) measurements \cite{wray10}. Nevertheless, a major discrepancy regarding the existence of Majorana zero modes (MZMs) seems to occur: while point-contact spectroscopy on {\cbi} shows a zero-bias conductance peak, suggestive of the presence of MZMs \cite{sasak11}, Andreev reflection spectroscopy and STM measurements demonstrate fully-gapped superconductivity with no in-gap states, indicative of the absence of MZMs \cite{peng13,levy13}. 
Such radically different outcomes from these two surface-sensitive techniques raise serious questions about the effective realization of topological superconductivity in {\mbi}. 

Recent experimental observations, however, support the unconventional pairing state associated with topological superconductivity. Several bulk probes, including nuclear magnetic resonance (NMR) \cite{matan16}, field-angle-resolved resistivity \cite{pan16,smyli18}, and heat-capacity measurements \cite{yonez17}, indicate a twofold anisotropy in the basal plane, in spite of the three-fold $D_{3h}$ symmetry of the lattice. Besides the above bulk measurements, also surface sensitive scanning tunneling microscopy (STM) measurements reveal a superconducting gap with a twofold symmetry, pinned by a mirror plane for in-plane magnetic fields \cite{tao18}. The observation of a  superconducting gap with twofold anisotropy indicates an odd-parity superconductivity which breaks the rotational symmetry of the underlying lattice, or nematic superconductivity. The latter is an analogue of liquid crystals, known to break the rotational symmetry while preserving translational symmetry \cite{fradk10}. The systematic investigation of 
nematic superconductivity is a pressing demand for a comprehensive understanding of topological superconductivity.

Recently, the binary stannide semimetal {\ca}, crystallizing in 
the cubic {\ac3}-type structure with point group {\oh} [Fig. 1(a) inset], has been proposed as a prime candidate for realizing topological superconductivity \cite{gupta17}. This stoichiometric compound undergoes a superconducting transition at {\tc} = 4.2\,K \cite{luo15a}, and is predicted to be a topologically non-trivial semimetal by theoretical calculations \cite{gupta17}. In the absence of SOC, its electronic structure is predicted to host topological nodal lines. 
Upon introducing SOC, the nodal lines evolve into topological point nodes. The non-trivial Berry phase associated with the topological nature of {\ca} is experimentally confirmed by recent quantum oscillation measurements \cite{zhu19,siddi21}. Moreover, the Fermi surfaces determined by such measurements are shown to enclose an odd number of TRIM \cite{siddi21}, thus fulfilling one of the two prerequisites for topological superconductivity. Clearly, it is essential to investigate whether {\ca} satisfies also the other prerequisite 
associated with the superconducting pairing states, namely, if it shows odd-parity pairing.

\begin{figure}[t]
\includegraphics[width=8cm]{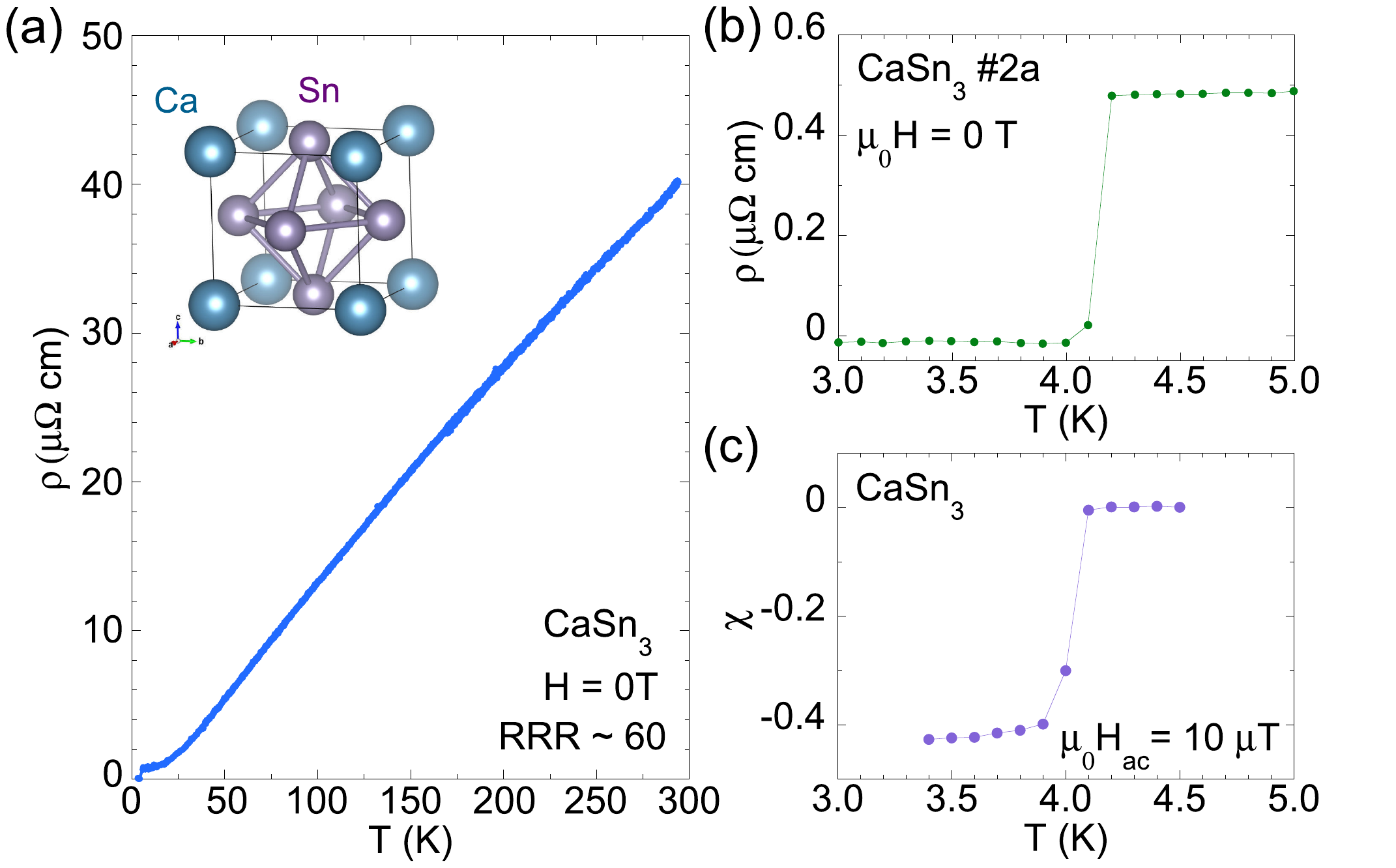}
\caption{\label{fig:res}(a) Inset: {\ca} crystallizes in the {\ac3}-type cubic 
structure with a fourfold symmetry around the $\langle$100$\rangle$-axes \cite{momma11}. Main panel: Temperature dependence of resistivity showing a high residual resistivity ratio $RRR \sim 60$. (b) Low-temperature resistivity of {\ca} ($T < 6$\,K), showing a superconducting transition at $T_c = 4.15$\,K, consistent with the previously reported value \cite{luo15a}. (c) ac magnetic susceptibility of {\ca} \cite{Nfactor}.}
\end{figure}

Here, we report on a detailed study of the superconducting state of the topological semimetal {\ca}. The measured upper critical fields $H_{c2}$ show a remarkable twofold anisotropy around a $C_{4}$ axis (in a crystal structure with point group {\oh}). Such twofold anisotropy in the SC state cannot be attributed to an anisotropic effective mass, nor to flux-flow depinning due to the Lorentz force. Together with the peculiar temperature dependence of $H_{c2}$, the observation of spontaneous rotational symmetry breaking in the superconducting state for {\ca} clearly evidences the realization of nematic superconductivity. Muon-spin relaxation and rotation (\musr) measurements indicate a nodeless superconducting pairing state without time-reversal-symmetry breaking. Moreover, {\musr} measurements in magnetic fields applied along two mutually orthogonal crystal directions evidence very different electron behaviors, thus supporting nematic superconductivity in {\ca}. Along with a previous study, which confirms an odd number of TRIM  enclosed by the Fermi surfaces of {\ca}, our results suggest that {\ca} is a promising nematic superconductor candidate.

\section{Experimental Methods}

Single crystals of {\ca} were grown using Sn self flux. The starting materials with a ratio of Ca:Sn = 1:9 were placed in an alumina crucible and sealed in a quartz tube. The quartz ampoule was heated up to 800 $^{\circ}$C, kept for 24 hours, and cooled slowly down to 300 $^{\circ}$C at a rate of 2 $^{\circ}$C/h, where the excess of molten Sn was decanted by centrifugation. Powder x-ray diffraction (XRD) results, shown in Fig. \ref{fig:xrd}a, confirm the {\ac3}-type cubic structure with a lattice constant $a = 4.7331(5)$\,{\AA}, consistent with previous work \cite{luo15a}. We also confirm a stoichiometry of Ca:Sn = 1:2.98 with x-ray fluorescence spectroscopy, suggesting that no excess of Sn flux inclusions remain in the final samples. As reported previously \cite{luo15a,zhu19}, {\ca} is very air-sensitive and decomposes into Sn on the surface. To avoid decomposition, we kept the crystals 
in a N$_{2}$-filled glove box and minimized exposure to air while conducting magnetic and transport measurements. Cut from the same starting piece, bar-shaped samples \#2a and \#2b ($\sim$ 1.5$\times$0.50$\times$0.075 mm$^{3}$) were used for the angle-dependent magneto-transport measurements with a four-wire configuration in a dilution refrigerator with a Swedish rotator. 
Depending on the temperature range, the applied electrical currents were up to 312~$\mu$A.
We aligned carefully the applied current direction to the crystal axis determined from the cubic facets, so that the applied magnetic fields could rotate in a crystallographic plane within a few degrees of misalignment.
We measured the ac susceptibility using a conventional mutual-inductance 
method with an ac driving field of 10~$\mu$T and a driving frequency of 991~Hz.

The \musr\ measurements were performed at the Dolly spectrometer ($\pi$E1 beamline) of the Swiss Muon Source at the Paul Scherrer Institute, Villigen, Switzerland. Experiments down to 0.27\,K 
were carried out by using a $^{3}$He cryostat. A mosaic of {\ca} single crystals was glued on a specially manufactured Ag sample holder. The full experimental details are reported in App.~\ref{app:A}. The incident muon momentum was parallel to the crystal [110] direction with the external magnetic field $\boldsymbol{H}$ applied parallel ($\mu_{0}H_{\parallel} \neq 0$) or orthogonal ($\mu_{0}H_{\bot} \neq 0$) to the [110] direction. The obtained results confirm that no degradation occurred to the investigated samples (see App.~\ref{app:B}).

\begin{figure*}[tbh]
\includegraphics[width=17cm]{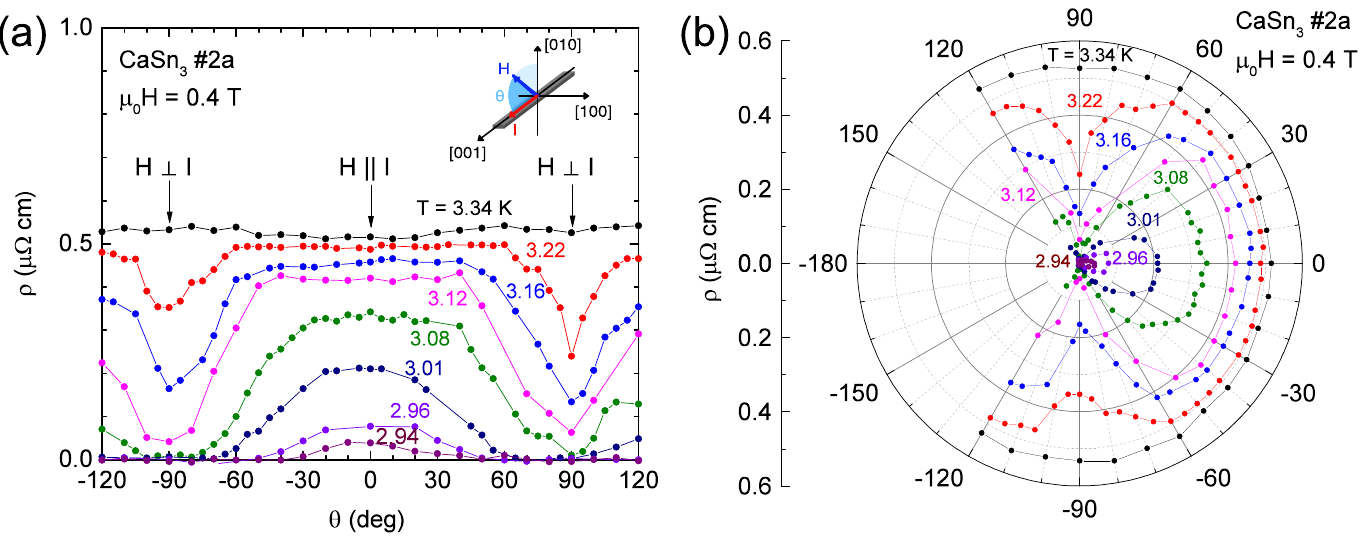}
\caption{\label{fig:RSB}(a) Resistivity as a function of $\theta$ for {\ca} at $\mu_{0}H$ = 0.4\,T at several temperatures, where $\theta$ is the angle between the applied field $H$ and the applied current $I\parallel [001]$. The upper critical field at $\theta=\pm 90^{\circ}$ ($H\perp I$) is higher than that at $\theta= 0^{\circ}$ ($H\parallel I$). (b) The polar plot highlights the prominent rotational-symmetry 
breaking in the upper critical field of {\ca}.}
\end{figure*}

\section{Experimental Results}
\subsection{Resistivity Measurements}
Upon lowering the temperature, the resistivity of {\ca} shows a metallic behavior, followed by a sharp superconducting transition at $T_{c}=4.15$\,K, close to the previously 
reported value (Figs.~\ref{fig:res}a and \ref{fig:res}b) \cite{luo15a}. The residual 
resistivity $\rho_{0}\sim 0.5$\,$\mu\mathrm{\Omega}$cm, measured right above $T_{c}$, is $1000$ times lower than that of the doped topological insulator {\mbi} \cite{pan16}. This result and the high residual resistivity ratio $RRR \sim 60$ suggest that the tested crystal has a low intrinsic crystalline disorder.


\subsection{Magnetic Susceptibility}
\label{ssec:magn_susc}

Despite a previous study \cite{zhu19} reporting intrinsic superconductivity below 1.2 K and surface superconductivity at 4 K (from surface decomposed Sn), here we bring unambiguous evidence of intrinsic superconductivity in CaSn$_3$ at $\sim 4.15$\,K by means of ac magnetic susceptibility. In our case, we finely ground a {\ca} single crystal and sealed it, along with two Cu wires acting as a thermal link, in a Kapton tube with Devcon rapid epoxy (see inset of Fig.~\ref{fig:xrd}a in App.~\ref{app:C}). The procedure was carried out in a glove box, thus excluding any possible exposure to air. An x-ray diffraction (XRD) pattern recorded immediately before the magnetic susceptibility measurements revealed \textit{no traces of Sn} within the threshold sensitivity of the XRD technique ($\sim$~2\%), as reported in App.~\ref{app:C}. 
In Fig.~\ref{fig:res}c, we show the ac magnetic susceptibility data, 
highlighting the onset of superconductivity at 4.15\,K, consistent with the zero-resistance temperature shown in Fig. \ref{fig:res}b. In our case, the sharpness of the transition and its onset $\sim 0.5$\,K above that of tin $T_{c}=3.72$\,K~\cite{Pool} definitively rule out the presence of surface-decomposed Sn and/or of Sn inclusions, originating from the self-flux synthesis technique. A lower bound for the shielding fraction of our sample is estimated to be slightly above 40\%~\cite{Nfactor}. Note that powdered samples generally show a reduced shielding due to the presence of a non-negligible fraction of grains whose size is of the same order of magnitude as the effective magnetic penetration depth. The \tfmu\ results reported in Sec.~\ref{ssec:musr} definitively 
demonstrate the bulk character of superconductivity in {\ca}.


\subsection{Upper Critical Fields}

We performed detailed angle-dependent magnetoresistance measurements at different magnetic fields on a set of {\ca} single crystals. Figure~\ref{fig:RSB}a shows the angle-dependent resistivity $\rho(\theta)$, where $\theta$ is the angle between the electric current $I\parallel [001]$ and the applied magnetic field $H$. Since the {\ac3}-type cubic structure with the point group $O_{h}$ has a four-fold rotational symmetry about the $\langle$100$\rangle$-axes, i.e., the $C_{4}$ axes, we expect a four-fold modulation of the physical quantities, including here the upper critical field.
However, the experimental data at $\mu_{0}H = 0.4$\,T clearly show only a twofold anisotropy. It is worth noting that no distinct anisotropy of $\rho(\theta)$ is observed in the normal state 
at $T = 3.34$\,K, instead, it manifests itself 
only below $T_{c}(0.4\,\mathrm{T})$. For instance, while being resistive at $\theta=0^{\circ}$ ($H\parallel I \parallel [001]$) at $T=3.08$ K, the sample is superconducting around $\theta=\pm 90^{\circ}$ ($H\parallel [010]\perp I$), indicating that $H_{c2}$ at $\theta=\pm 90^{\circ}$ is notably higher than that at $\theta=0^{\circ}$. The spontaneous rotational symmetry breaking is even more evident in the polar plot of $\rho(\theta)$ shown 
in Fig.~\ref{fig:RSB}b.
The observed angular dependence of $H_{c2}$ can be ascribed to several effects, 
as described in detail below.

\begin{figure}[t]
\includegraphics[width=8cm]{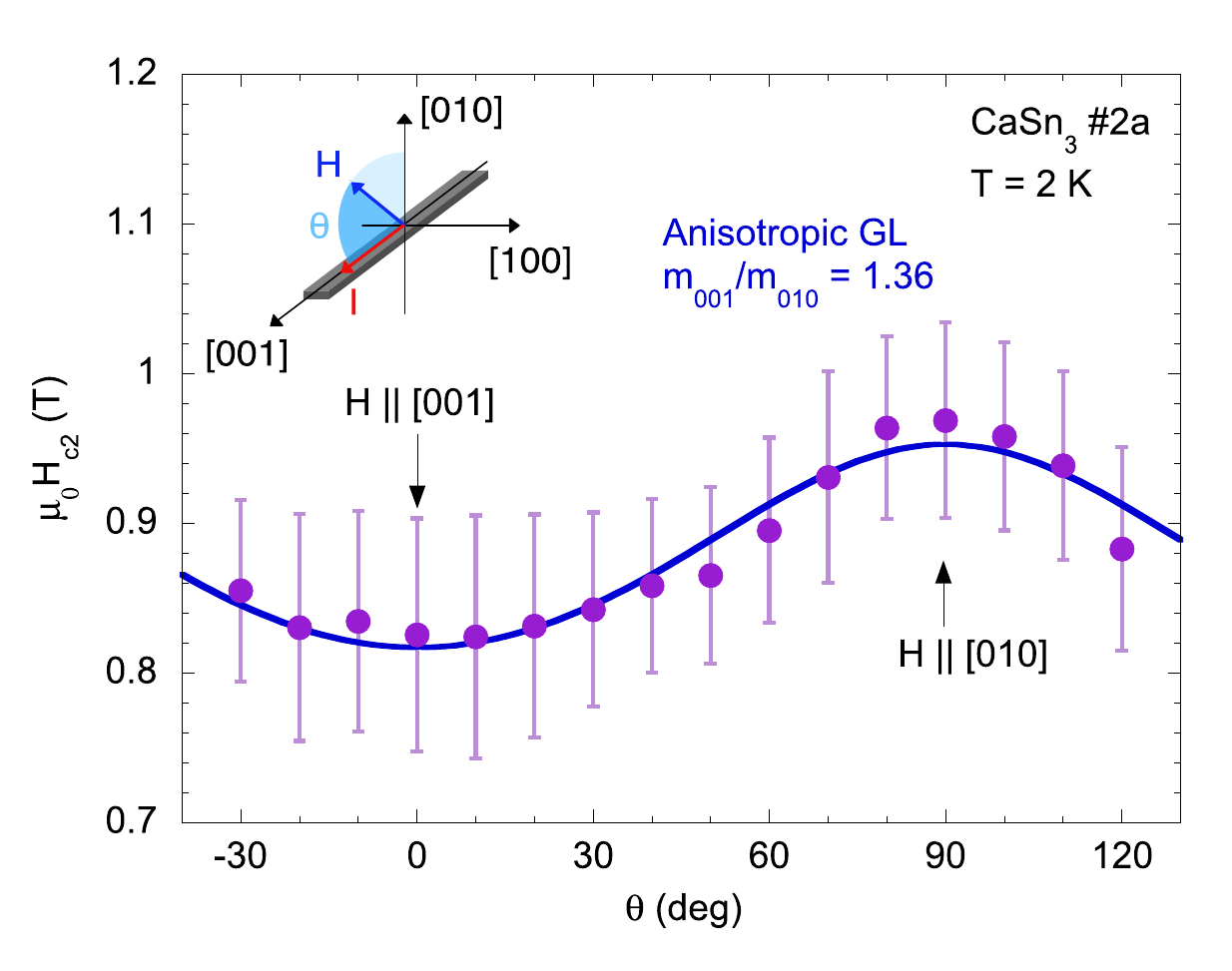}
\caption{\label{fig:Hc2_aniso}Upper critical field $H_{c2}$ obtained from the midpoints of resistive transitions as a function of $\theta$. The error bars indicate $H_{c2}$ determined by 10\%-90\% resistive transitions. The blue line is the best fit to data using the anisotropic GL model, $H_{c2}(\theta)=H_{c2}(0^{\circ})/\sqrt{\cos^{2}\theta + \Gamma^{-2}\sin^{2}\theta}$, with extracted effective mass anisotropy $\Gamma^{2} = (H_{c2}(90^{\circ})/H_{c2}(0^{\circ}))^{2}=m_{001}/m_{010}$ of 1.36 $\pm$ 0.03 and $\mu_{0}H_{c2}(0^{\circ})$ = 0.82 $\pm$ 0.01 T. }
\end{figure}

First of all, in the framework of the anisotropic Ginzburg-Landau (GL) model, the angular behavior of $H_{c2}$ reflects the anisotropy of the effective mass. In this case, the angular dependence of the upper critical field is described by:
\begin{equation}
  H_{c2}(\theta) =\frac{H_{c2}(0^{\circ})}{\sqrt{\cos^{2}\theta + \Gamma^{-2}\sin^{2}\theta}},
\end{equation}
where the anisotropy ratio $\Gamma$ is given by $\Gamma = H_{c2}(90^{\circ})/H_{c2}(0^{\circ}) = \sqrt{m_{001}/m_{010}}$. Here $m_{001}$ and $m_{010}$ are the effective masses for the energy dispersion along the [001] and [010] directions, respectively. The extracted $\Gamma$ from a fit to the measured angular variation of $H_{c2}(\theta)$ is 1.17, yielding an unusually large mass anisotropy of $\Gamma^{2} = m_{001}/m_{010}\sim 1.36$ (Fig.~\ref{fig:Hc2_aniso}). However, in cubic systems with a point group {\oh}, the effective masses along the main orthogonal axes are isotropic, i.e., $m_{001} = m_{010}$, incompatible with the values obtained for {\ca}, thus excluding that the twofold oscillations in {\hc2} can arise from the anisotropy of the effective mass.

\begin{figure*}[tb]
\includegraphics[width=17cm]{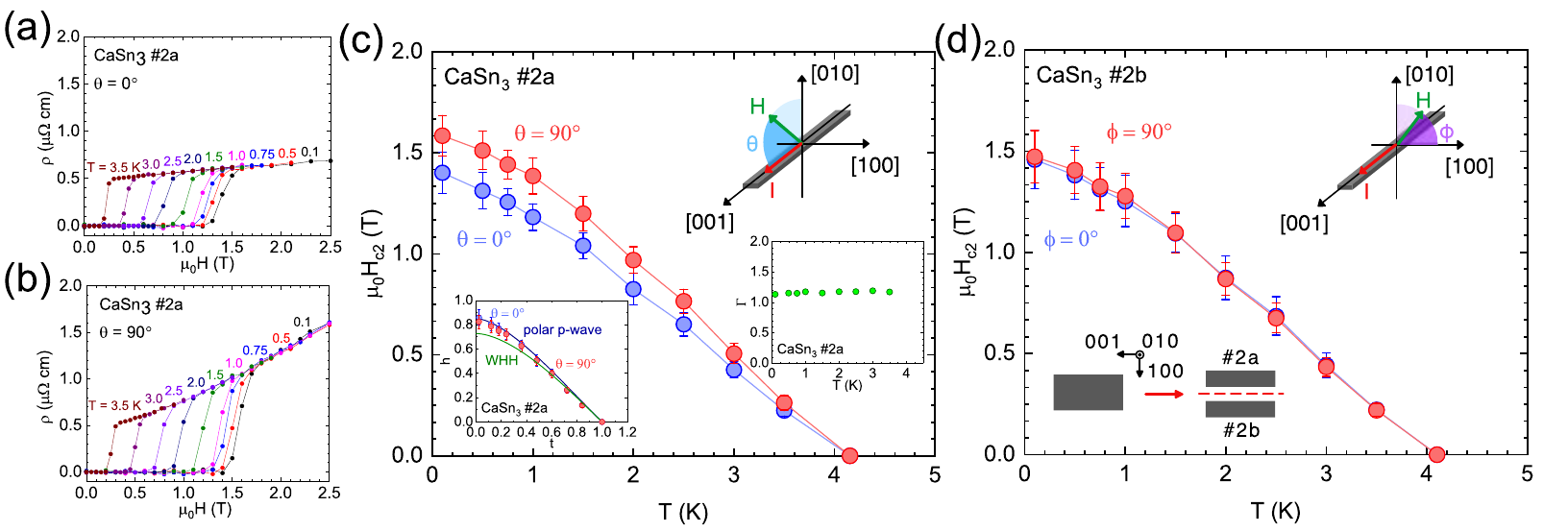}
\caption{\label{fig:Hc2}Resistivity as a function of magnetic field in the configurations: (a) $H\parallel I$ ($\theta = 0^{\circ}$) and (b) $H\perp I$  ($\theta = 90^{\circ}$) for {\ca} (sample \#2a). Here, $\theta$ is the angle between the applied field $H$ and the current $I\parallel [001]$ (upper inset). A large magnetoresistance is observed in the $H\perp I$ configuration due to the orbital effect. (c)  Temperature dependence of the upper critical field of the sample \#2a for $\theta = 0^{\circ}$ and $\theta = 90^{\circ}$, obtained from the midpoint of resistive transitions with error bars indicating 10\%-90\% values plotted in (a) and (b). A clear breaking of the rotational symmetry is observed. The lower left inset shows the normalized upper critical field $h=-H_{c2}/(T_{c}\left. \frac{dH_{c2}}{dT}\right |_{T=T_{c}})$ vs the normalized temperature $t=T/T_{c}$. Solid lines are theoretical curves based on the WHH theory (green) and on calculations for a polar $p$-wave pairing state (blue) \cite{schar80}. The lower right inset depicts the anisotropy $\Gamma= H_{c2}(90^{\circ})/H_{c2}(0^{\circ})$ as a function of temperature. (d) Temperature dependence of the upper critical field of the sample \#2b for $\phi = 0^{\circ}$ and $\phi = 90^{\circ}$, where $\phi$ is the angle between $H$ and the [100] axis (upper inset). The lower inset illustrates the samples \#2a and \#2b, cut from the same starting piece.}
\end{figure*}

Another possible explanation for the measured $H_{c2}$ anisotropy in
{\ca} is a lowering of symmetry due to flux-flow resistance. 
Since the Lorentz force depends on the relative angle between $H$ and $I$, flux lines in type-II superconductors can be depinned from pinning centers. In the longitudinal configuration $(H\parallel I)$, no Lorentz force is produced by the applied magnetic field. By contrast, 
in the transverse configurations $(H\perp I)$, the applied field 
induces a Lorentz force, leading to a finite flux\-flow resistance that reduces $H_{c2}$, accompanied by a broadening of the superconducting transition. The large orbital magnetoresistance caused by the Lorentz force in the normal state of {\ca} in the transverse configuration (Fig.~\ref{fig:Hc2}b), $H_{c2}(90^{\circ})$ ($H\perp I$), is higher than $H_{c2}(0^{\circ})$ ($H\parallel I$), inconsistent with the flux-flow depinning hypothesis. Together with the absence of broadening of the resistive transition in a transverse configuration (see Figs.~\ref{fig:Hc2}a and b), this 
excludes the flux-flow scenario as a cause for the observed anisotropy. We note that the observed anisotropy $(H_{c2}\perp I>H_{c2}\parallel I)$ also disagrees with the anisotropy caused by Sn potentially decomposed on the surface, as this would exhibit an opposite anisotropy $(H_{c2}\parallel I>H_{c2}\perp I)$ \cite{harpe68}.

Last but not least, the unusual $H_{c2}$ angular behavior could, in principle, originate from the lowering of the crystallographic symmetry due to a low-temperature structural phase transition. However, we did not observe any discernible anomaly --- normally associated with such phase transitions --- either in the temperature dependence of resistivity (Fig.~\ref{fig:res}), or in that of heat capacity \cite{luo15a}. In addition, the temperature-independent anisotropy of {\hc2} rules out a possible structural phase transition at or indistinguishably close to $T_{c}$. If a structural phase transition occurs at {\tc}, the anisotropy of {\hc2} should show a steep increase below {\tc}, mimicking the temperature dependence of the structural order parameter, here closely linked to the lattice distortion \cite{axe73}. However, this is irreconcilable with the observed temperature dependence of {\hc2} anisotropy in {\ca} (lower right inset in Fig.~\ref{fig:Hc2}c), suggesting that the crystallographic symmetry is preserved, i.e., it retains the same point group {\oh}, also at low temperatures. In addition, as explained later on, the twofold anisotropy of $H_{c2}$ does not stem from a misalignment of the crystallographic axes either.

Once the GL anisotropic model and the flux-flow scenario are excluded, we can attribute the unexpected rotational symmetry breaking in the superconducting state of {\ca} to an unconventional pairing symmetry. Indeed, a very similar spontaneous rotational symmetry breaking in the superconducting state, or nematic superconductivity, 
was previously observed in {\mbi} (a putative TSC) via field-angle-resolved resistivity~\cite{pan16,smyli18}. Here, the nematic behavior involves the odd-parity pairing symmetry 
described by a two-dimensional $E_{u}$ representation in the trigonal 
$D_{3d}$ point group of {\mbi}~\cite{vende16}.

In {\ca}, in addition to the nematic behavior, the peculiar temperature dependence of $H_{c2}$ supports the realization of unconventional superconducting pairing. Regardless of the applied-field configuration, $H\parallel I$ or $H\perp I$, the {\hc2} of {\ca} extracted from the midpoints of resistive transitions (Figs.~\ref{fig:Hc2}a and b) increases linearly with decreasing temperatures down to $T\sim 0.4 T_{c}$, deviating from the conventional orbital depairing field described by the Werthamer-Helfand-Hohenberg (WHH) theory \cite{werth66,helfa66}. According to the WHH theory, $H_{c2}(T)$ at $T=0$\,K is given by $H_{c2}(0)=-\alpha T_{c}\left.\frac{dH_{c2}}{dT}\right |_{T=T_{c}}$, where $\alpha$ is 0.69 in the dirty limit and 0.73 in the clean-limit superconductors (Fig.~\ref{fig:Hc2}c)~\cite{werth66,helfa66}. However, the obtained values of $H_{c2}(0) = 1.4$\,T for $\theta =0^{\circ}$ ($H\parallel I \parallel [001]$) and 1.6\,T for $\theta = 90^{\circ}$ ($H\parallel [010] \perp I$) correspond to $\alpha\sim 0.85$, independent of $\theta$, substantially exceeding the orbital limit for clean superconductors $\alpha= 0.73$ (Fig.~\ref{fig:Hc2}c inset). The overcoming of the orbital limit can be ascribed to several reasons, such as Fermi-surface topology \cite{kita04,shiba06} or multiband effects \cite{schar80}. However, an enhancement of {\hc2} due to Fermi-surface topology would reflect the anisotropy of the crystal structure. In this case, the anisotropy of {\hc2} would show a fourfold symmetry about the $C_{4}$ axes, at odds with the observed twofold oscillations. The multiband effect is also implausible, because the measured {\hc2} lacks the peculiar positive curvature of the temperature dependence of multiband superconductors, such as MgB$_{2}$~\cite{gurev04}. Instead, {\hc2} of {\ca} is in a good agreement with that of a polar $p$-wave pairing state~\cite{schar80}, as shown in the inset of Fig.~\ref{fig:Hc2}c, where we plot the normalized upper critical field $h=-H_{c2}/(T_{c}\left. \frac{dH_{c2}}{dT}\right |_{T=T_{c}})$. The good agreement between theory (blue line) and data suggests the realization of an unconventional pairing SC state.

Finally, it is important to note that upon rotating the magnetic field in the $C_{4}$ plane perpendicular to the applied current ($\phi$ rotation, from [100] to [010]), we observe no traces of twofold anisotropy: $H_{c2}(\phi =0^{\circ})\parallel [100]$ fully overlaps with $H_{c2}(\phi =90^{\circ})\parallel [010]$, as shown in Fig.~\ref{fig:Hc2}d. This effect cannot be ascribed to a trivial misalignment of the current direction with respect to the crystallographic axes, otherwise we would have found traces of a two-fold anisotropy, clearly not the case here. This finding indirectly confirms that the twofold anisotropy upon $\theta$ rotation is due to unconventional pairing. 
At the same time, it suggests that, in {\ca}, the nematic director is oriented along a crystallographic axis, i.e., [001], presumably because of the formation of a single domain or the imbalance in multiple domains of the nematic state due to uniaxial strain fields (such as a slight lattice distortion). Evaluated from the GL relations, the superconducting coherence lengths in the [100], [010], and [001] directions are $\xi_{100}=\xi_{010}=153$\,{\AA} and $\xi_{001}=133$\,{\AA}, implying anisotropic superconducting pairing interactions due to nematicity.  
In {\mbi}, the nematic director does not depend on the presence or absence of an applied current \cite{matan16,yonez17,smyli18,kunts18,tao18}, but it is correlated with tiny distortions of the hexagonal lattice $(\Delta a/a \sim 0.02 \%)$ \cite{kunts18}. In {\ca}, the upper limit of structural distortion --- as estimated from peak broadening in XRD measurements --- is $\Delta a/a \sim 0.06$ \%, most likely associated with a pinning of the nematic director.


\subsection{\label{ssec:musr}{\musr} measurements}

\begin{figure}[tbh]
\includegraphics[width=0.45\textwidth]{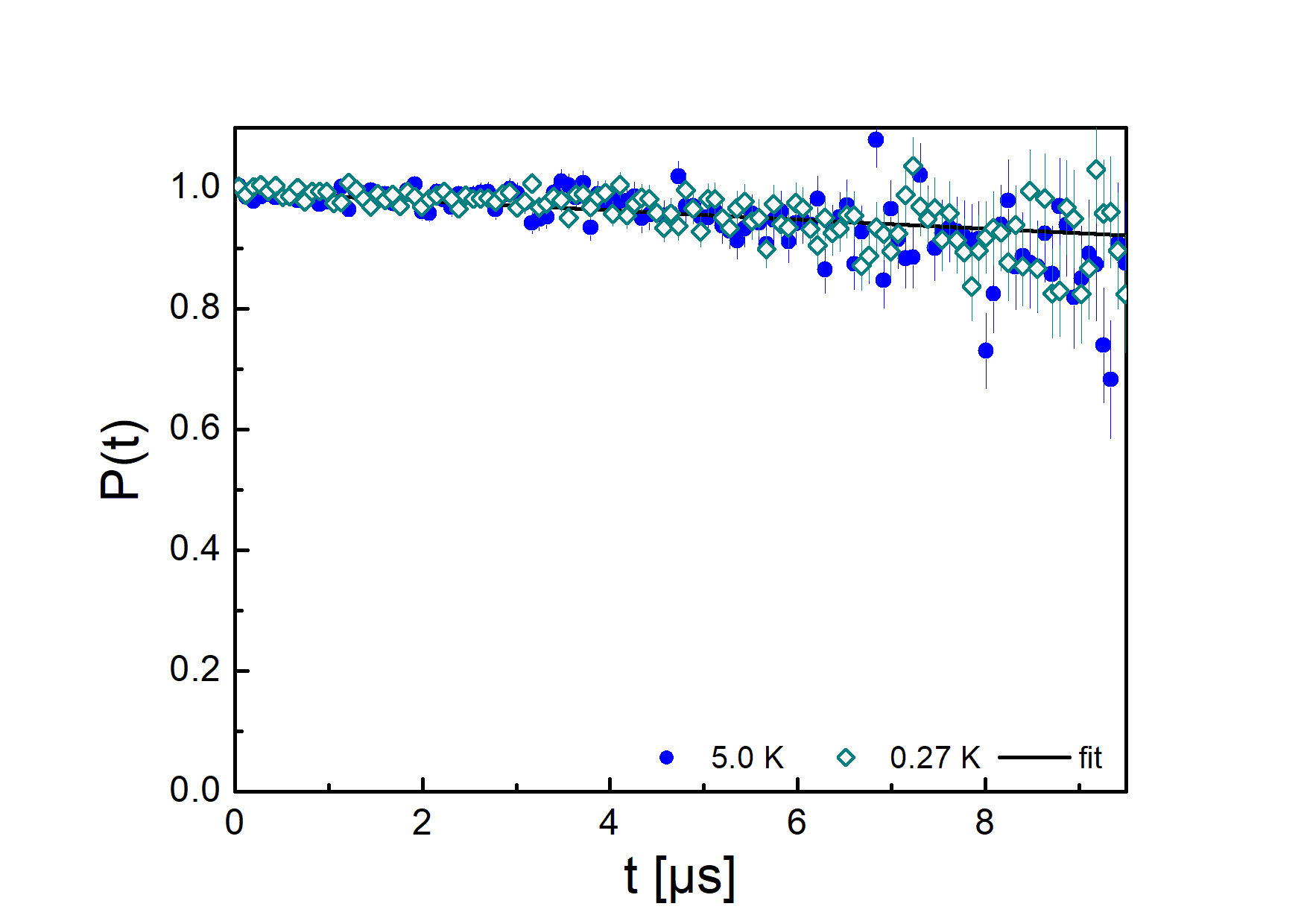}
\caption{\label{fig:ZF}Time-domain zero-field muon-spin polarizations 
at 0.27\,K (open symbols) and at 5\,K (full symbols). The line represents 
a simple exponential fit to the basically overlapping datasets. 
See text for details.}
\end{figure}

The true nature of the topological superconducting state and of its nematic properties are often controversial. To address this important issue in the {\ca} case, we performed systematic measurements by means of muon spectroscopy {\musr}, an extremely sensitive bulk probe of the local (i.e., microscopic) electronic properties. This technique uses spin-polarized positive muons implanted in the sample under test \cite{Blundell1999,Amato1997}. Depending on the material density, $\mu^{+}$ typically penetrate over a depth of several hundreds of microns and, thus, are implanted homogeneously over the whole sample volume. Consequently, muons can be considered as a bulk probe of the local electronic properties.
 
\begin{figure*}[tbh]
\includegraphics[width=\textwidth]{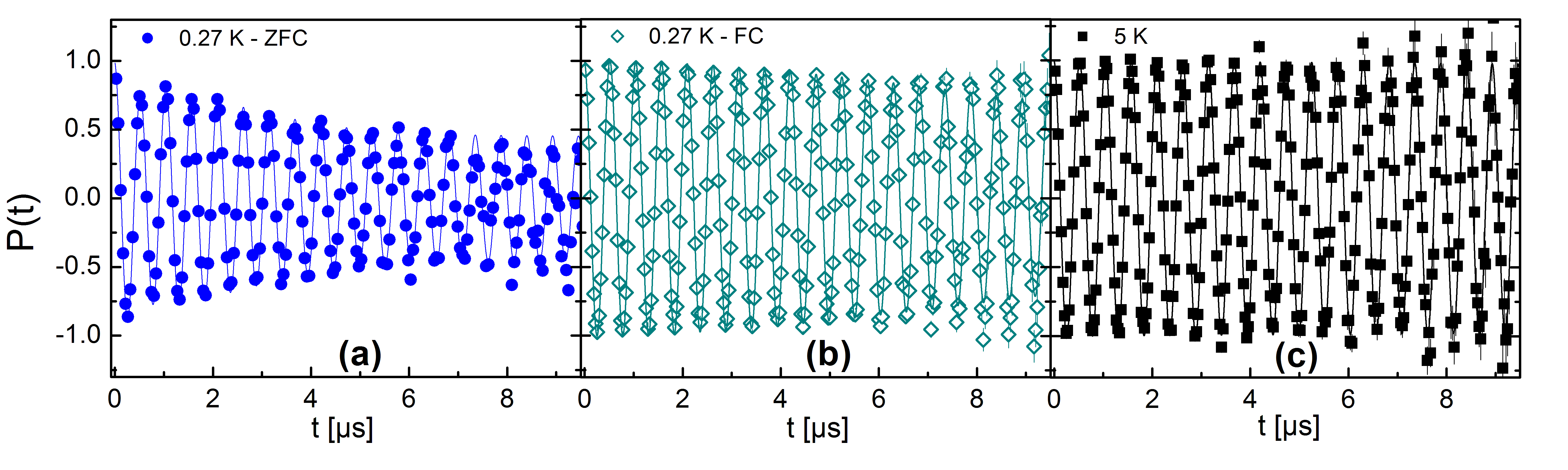}
\caption{\label{fig:TFmuSR}Muon-spin polarization $P(t)$ in an 
applied field of $\mu_{0}H_{\perp}= 14$\,mT in the [001] crystal direction at 0.27\,K after ZFC (a), after FC (b), and at 5\,K (c). Solid lines are fits to the model described by Eq.~(\ref{eq:pt}).}
\end{figure*}

We performed two kind of experiments, (i) zero-field muon spectroscopy ({\zfmu}) and (ii) transverse-field muon spectroscopy ({\tfmu}) measurements. As to the former, we searched for the presence of possible tiny local magnetic fields, typical of, e.g., superconductors with spin-triplet pairing, as observed in Sr$_2$RuO$_4$ \cite{Luke1998,Shiroka2012} or in TSCs-like Sr$_{0.1}$Bi$_2$Se$_3$ \cite{Biswas2019}.
In Fig.~\ref{fig:ZF} we present the ZF time-dependent polarization below and above the superconducting transition. In both cases the experimental data are well fitted by a \textit{purely} Gaussian decay, typically associated with the small static magnetic fields due to the randomly-oriented nuclear magnetic dipole moments on the muon implantation sites \cite{Yaouanc2011}. Both datasets are completely superposed, thus excluding a possible increase of the decay of polarization from spontaneous magnetic fields occurring in the superconducting state. This finding rules out the breaking of time-reversal symmetry in the superconducting state of {\ca}, thus excluding possible chiral superconducting states, such as $p_{x}\pm ip_{y}$ \cite{kalli16}.

We performed {\tfmu} measurements following both standard field-cooling (FC) and nonstandard zero-field cooling (ZFC) protocols. The latter was used to evaluate the superconducting volume fraction. To this aim, after cooling down to 0.27 K in ZFC mode, we applied a magnetic field $\mu_{0}H_{\perp}=$ 14 mT along the [001] direction, orthogonal to the incident muon spin ${\bf S}_{\mu}$(see App.~\ref{app:A}). The ZFC protocol 
produces rather inhomogeneous local magnetic fields at the implanted muon sites, reflecting an artificially disordered vortex lattice in the SC state. In turn, the disordered vortex lattice causes a much higher muon-spin depolarization rate (Fig.~\ref{fig:TFmuSR}a) than that obtained following a conventional FC procedure (Fig.~\ref{fig:TFmuSR}b) or in the normal state (Fig.~\ref{fig:TFmuSR}c). The time-dependent muon-spin polarization $P(t)$ represented in Fig.~\ref{fig:TFmuSR}a can be described by the following model:
\begin{equation}\label{eq:pt}
  \begin{split}
    P(t) =f_\mathrm{sc}\cos(\gamma_{\mu}B_{\mu}t+\phi)\exp\left(-\frac{\sigma_{n}^{2}+\sigma_\mathrm{sc}^{2}}{2}t^{2}\right)\\
    +f_\mathrm{tail}\cos(\gamma_{\mu}B_\mathrm{tail}t+\phi)\exp(-\Lambda t).
  \end{split}
\end{equation}

Here, $f_\mathrm{sc}=1-f_\mathrm{tail}$ is the volume fraction of those muons probing the vortex state and, hence, it represents the superconducting fraction. $f_\mathrm{tail}$ is the fraction of muons implanted in the silver sample-holder (and in the non superconducting part of the sample), $\gamma_{\mu}/2\pi= 135.53$ MHz/T is the muon gyromagnetic ratio, $B_{\mu}$ and $B_\mathrm{tail}$ are the magnetic fields probed by muons implanted in the sample and in the silver plate, $\phi$ is the initial phase. $\sigma_\mathrm{n}$ and $\sigma_\mathrm{sc}$ are the Gaussian depolarization rates in the normal and superconducting state, respectively. $\sigma_\mathrm{n}$ is due to the static nuclear dipolar fields and it is negligibly small in this compound. Note that, in our case, 
$\sigma_\mathrm{n}$ cannot be distinguished from the Lorentzian relaxation rate in silver. 
Therefore, it was fixed to zero and an eventual nuclear dipolar contribution was accounted for by the relaxation rate of muons implanted in the silver sample holder, $\Lambda$. Such parameter was obtained by fitting the data at 5\,K. The resulting value, $\Lambda = 0.005(2)~\mu$s$^{-1}$, was kept fixed in all the subsequent fit iterations. At the same time, the amplitude of the long-time tail of the ZFC {\tfmu} at 0.27\,K (Fig. \ref{fig:TFmuSR}a) is due to those muons implanted in the silver sample-holder (and in non superconducting parts of the sample). From the initial asymmetry amplitude and its long-time value we could estimate $f_\mathrm{tail}=A_\mathrm{tail}/A(0)$ and, therefore, determine a \emph{lower bound} for the superconducting volume fraction $f_\mathrm{sc} = 52\pm 1$\%, which confirms the bulk nature of superconductivity in this compound.
With the same applied field ($\mu_{0}H_{\perp}= 14$\,mT in the [001] direction) we performed a {\tfmu} 
temperature scan, after cooling the sample in field down to base temperature. Since such procedure ensures a homogeneous flux-line lattice, in principle, it allows us to determine the superfluid density $n_{s}$, the magnetic penetration depth $\lambda$, and their temperature dependence. In Fig. \ref{fig:TFmuSR}b, we show the time-dependent muon-spin polarization $P(t)$ at 0.27\,K. As expected, $P(t)$ decays more slowly than in the ZFC case (Fig. \ref{fig:TFmuSR}a). Nevertheless, such decay is still slightly faster than that measured above $T_{c}$ (shown here in Fig. \ref{fig:TFmuSR}c). The depolarization rate was then extracted by fitting the time-dependent polarization data to Eq. \eqref{eq:pt} while keeping $\Lambda$, $f_{Ag}$, and $f_\mathrm{sc}$ fixed at all temperatures. By assuming an ideal triangular vortex lattice, since the applied field is significantly lower than the measured {\hc2}, one can link $\sigma_\mathrm{sc}$ to the magnetic penetration depth $\lambda$ 
through the equation~\cite{Barford1988,Brandt2003}:
\begin{figure}[tbh]
\centering
\vspace{3mm}
\includegraphics[width=0.9\linewidth]{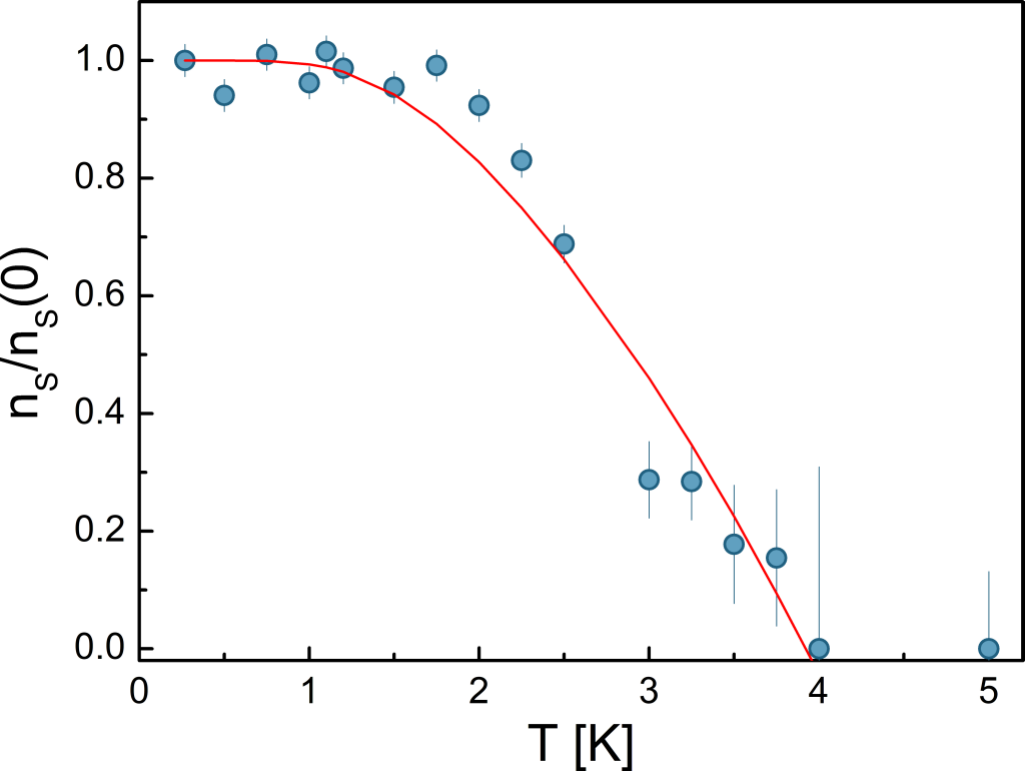}
\caption{\label{fig:ns}Normalized superfluid density $n_{s}(T)/n_{s}(T_{min})$ 
in an applied magnetic field $\mu_{0}H_{\perp}=$ 14\,mT along the [001] crystal 
direction. The solid line is a fit to Eq.~\eqref{eq:nsBCS}.}
\end{figure}
\begin{equation}\label{eq:lam}
  \frac{\sigma_\mathrm{sc}^{2}}{\gamma_{\mu}^{2}}=0.00371\frac{\Phi_{0}}{\lambda^{4}},
\end{equation}
with $\Phi_{0}$ the magnetic flux quantum. Since the magnetic field is 
applied along the [001] direction, Eq.~\eqref{eq:lam} 
allows us to determine the magnetic penetration depth due to the 
SC screening currents flowing in the (001) planes, i.e., 
$\lambda_{(001)} = 910(10)$\,nm at 0.27\,K. From the temperature 
dependence of the depolarization rate $\sigma_\mathrm{sc}$ in the 
superconducting state we can extract the superfluid density $n_{s}(T)$ by means of the relation between 
$n_{s}$ and $\sigma_\mathrm{sc}$:
\begin{equation}\label{eq:ns}
\frac{n_{s}(T)}{n_{s}(T_\mathrm{min})}=\frac{\lambda^{2}(T_\mathrm{min})}{\lambda^{2}(T)}=\frac{\sigma_\mathrm{sc}(T)}{\sigma_\mathrm{sc}(T_\mathrm{min})}.
\end{equation}
The normalized superfluid density $n_{s}(T)/n_{s}(T_\mathrm{min})$ of CaSn$_3$ is shown in Fig.~\ref{fig:ns}. Below $T_{c} \sim 4$\,K, it starts to increase with decreasing temperature, to saturate at low temperatures, below $\sim T_{c}/3$, suggestive of a nodeless superconducting state. Data were fitted by assuming a weak-coupling Bardeen-Cooper-Schrieffer (BCS) model for the superfluid density in the clean limit \cite{Tinkham}:
\begin{equation}\label{eq:nsBCS}
\frac{n_{s}(T)}{n_{s}(T_\mathrm{min})}=1-2 \int_{\Delta (0)} ^{\infty} \left( - \frac{\partial f }{ \partial E} \right) \cdot \frac{E dE}{\sqrt{E^{2}-\Delta^{2}}},
\end{equation}
where $\Delta(0)$ is the energy gap at zero temperature, while its temperature dependence is assumed to be $\Delta(T)/\Delta(0)=\tanh{1.82[1.018(T_{c}/T-1)]^{0.51}}$ \cite{Carrington2003,Rustem2008}.

\begin{figure}[tbh]
\centering
\vspace{-3mm}
\includegraphics[width=1.1\linewidth]{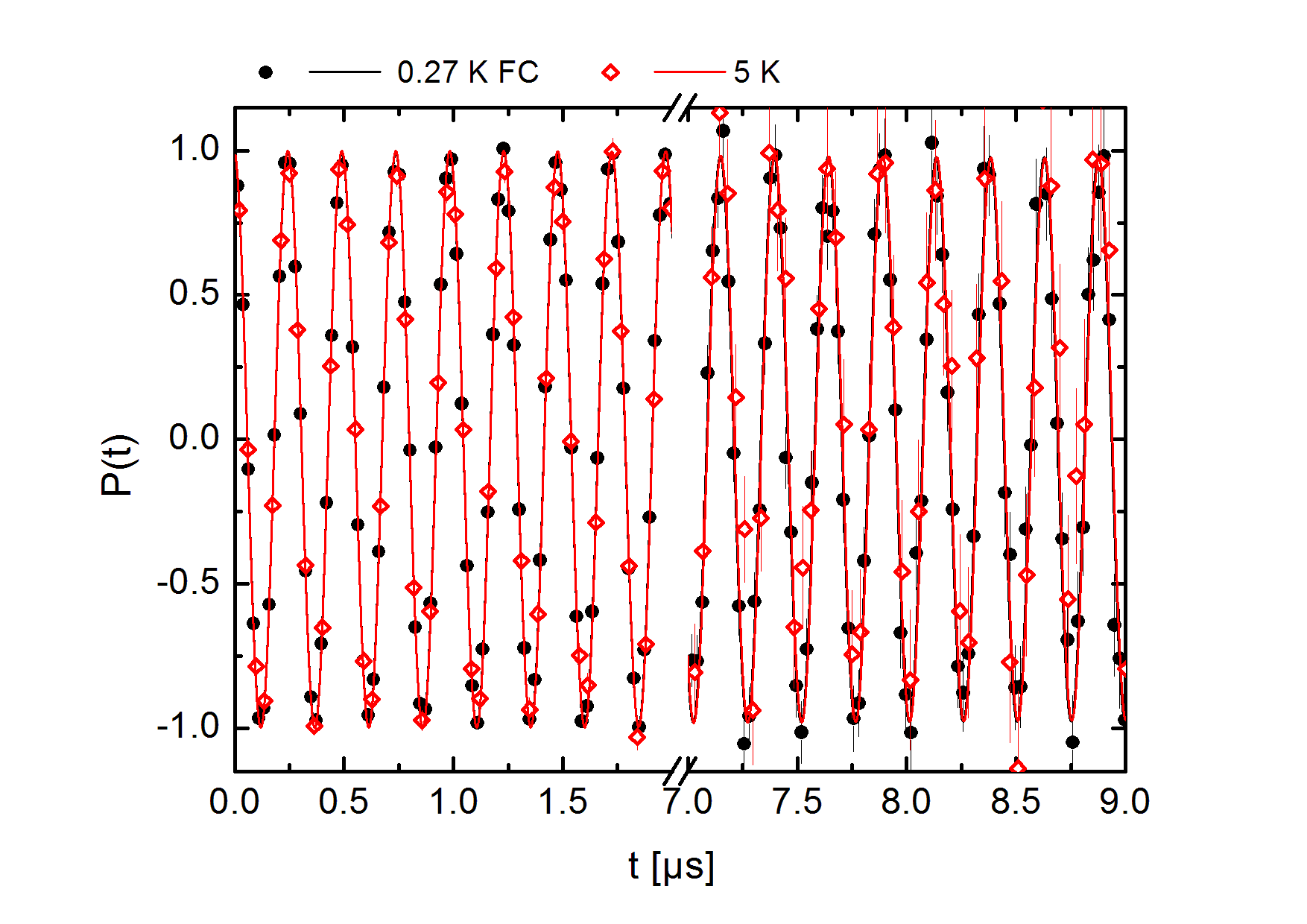}
\caption{\label{fig:para}Short-time and long-time TF muon-spin polarization recorded at 0.27 K (after FC) and at 5 K, in an applied magnetic field $\mu_{0}H_{\para}$ = 30 mT along the [110] crystal direction. The lack of a significant muon-spin depolarization below $T_c$ can be attributed to the presence of a non-trivial vortex phase (see text for details).}
\end{figure}

The numerical fit of the normalized superfluid density to Eq.~\eqref{eq:nsBCS} is in fair agreement with the experimental data except between 1.5 and 2.5\,K, where an unphysically saturated behavior still persists up to 2\,K. Beyond this temperature a rather sharp decrease occurs. This lack of sensitivity at intermediate temperatures could be due to the significantly low value of the measured depolarization rate. The parameters resulting from a numerical fit are $\Delta(0)\simeq 0.61(7)$\,meV and $T_{c} \simeq 3.9(2)$\,K. It is worth noting that, in {\ca}, both the pairing mechanism and the gap function most likely are far from those expected in standard BCS-like superconductors. Therefore, the reported values should be considered only as rough estimates.

To verify if the breakdown of rotational symmetry in {\ca}, observed via transport measurements through the twofold anisotropy of $H_{c2}$, is reflected also in its magnetic properties, we performed {\tfmu} measurements by applying the magnetic field of $\mu_{0}H_{\para} = 30$\,mT in the [110] direction. Such field value is clearly in the $H_{c1}<H_{\parallel}<H_{c2}$ range, as shown by our $H_{c2}$ measurements and by data from Ref.~\onlinecite{luo15a}. This ensures that the sample is in its mixed superconducting state after a field cooling down to 0.27\,K. In this case, we expect an ordered SC vortex state, which should induce a measurable depolarization rate in the muon-spin asymmetry. However, as shown in Fig. \ref{fig:para}, we could not detect any additional depolarization effects below $T_{c}$: the sample behaves as if it were in the normal state even at base temperature, clearly at odds with a $\mu_{0}H_{c2}$ value of at least 1.4\,T at 0.2\,K (see Fig. \ref{fig:Hc2}). This unexpected behavior could be the fingerprint of a non-trivial vortex phase. Such unconventional picture has been predicted to occur in superconductors with a multi-component order parameter~\cite{sigri91,Zyuzin2017,How2020}. Therefore, the present finding represents additional evidence of nematicity in the superconducting state of {\ca}.

\section{Discussion}

\begin{table}[t]
\includegraphics[width=0.45\textwidth]{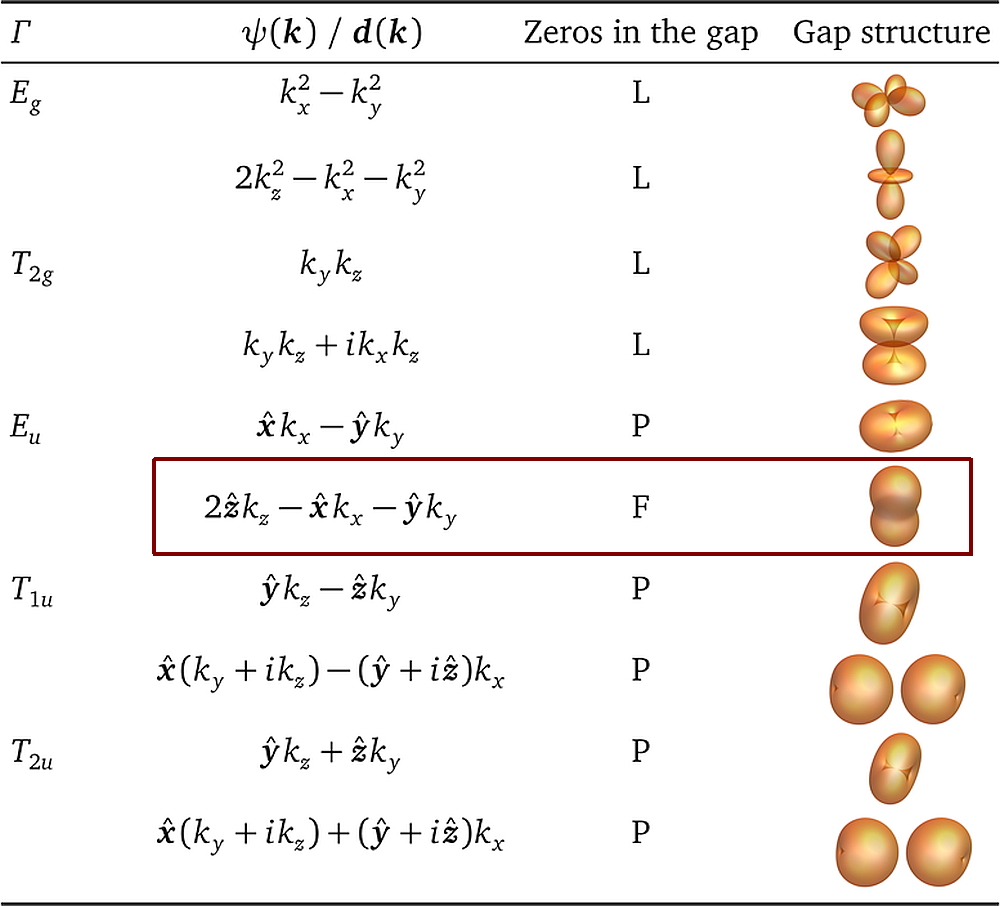}
\caption{\label{tab:GT}{\bf Nematic superconducting states with $D_{4h}$ symmetry for the point group $O_{h}$}. Gap function $\psi(\boldsymbol{k})$/$d$-vector $\boldsymbol{d}(\boldsymbol{k})$ for each of the $\Gamma$ representations with $D_{4h}$ symmetry. We also show the corresponding type of zeros in the gap of the quasiparticle excitation spectrum (P: point nodes, L: line nodes, F: full gap), as well as a graphical representation of the gap structure \cite{sigri91}. The highlighted row indicates the only fully-gapped odd-parity pairing state compatible with the experimental data.}
\end{table}

Angle-resolved magnetotransport measurements in {\ca} show that its $H_{c2}$ exhibits a twofold symmetry 
about a $C_{4}$ axis, an anisotropic behavior that breaks the rotational symmetry of the underlying crystal lattice. 
We ascribe this twofold anisotropy to a nematic director oriented along a crystallographic axis, as e.g., [001]. The same breaking of rotational symmetry was confirmed also by \musr\ measurements, which indicate a null magnetic response when the magnetic field is applied along the [110] direction, while showing a maximum response for a field applied parallel to the [001] direction. In this case, several qualitative considerations can be drawn.

First, it is particularly relevant to mention the highly unusual mixed state of \cbi, as evidenced by 
STM measurements~\cite{tao18}. Here, fields applied perpendicular 
to the selenium planes induce a pseudo-hexagonal vortex lattice. 
Interestingly, each SC vortex appears to be elongated in one direction, 
thus exhibiting an ellipsoidal shape. 
When, instead, the field was applied parallel to the Se planes, 
\emph{no obvious evidence of SC vortices} could be found, despite the 
sample being in the mixed superconducting state and the gap succesfully measured also in such configuration. This finding was justified by noting that, in a TSC, superconductivity on the surface is more robust against orbital depairing from an in-plane field, implying that the bulk vortex lines are pushed away from the surface~\cite{tao18}. Such scenario is inconsistent with our TF-measurements: contrary to STM, \musr\ spectroscopy is a bulk technique, thus insensitive to surface states. Moreover, the absence of vortices would mean that the lower critical field in the bulk is higher than the applied field, thus resulting in a full diamagnetic effect. In such case, the implanted muons do not sense any local field, except for the small nuclear dipolar fields, almost negligible in CaSn$_3$. Hence, a sizable non oscillating component in muon spin polarization and a corresponding peak at $B_{\mu}=0$ in the FFT would have been detected, at odds with our experimental results (see Fig.~\ref{fig:para} 
and App.~\ref{app:B}). 
Conversely, large and deformed vortices, far off the scenario of conventional Abrikosov SC vortices, 
can give rise to unexpectedly low modulations of the local magnetic field at the muon implantation sites and, consequently, to low muon depolarization rates for applied fields \emph{parallel} to the [001] direction, or even to a negligible modulation for applied fields \emph{parallel} to the [110] direction. 

Secondly, we find a mismatch between the carrier densities in the superconducting- $n_{s}$ and in the normal state $n_{n}$ of {\ca}. Our recent de Haas-van Alphen (dHvA) measurements provide an effective mass $m^{\ast}=0.18m_{e}$ and a carrier density $n_{n}=9.3\times 10^{20}$\,cm$^{-3}$ for the predominant hole band located at the $\Gamma$ point \cite{siddi21}. On the other hand, utilizing $\lambda_{0}=\sqrt{m^{\ast}/\mu_{0}n_{s}e^{2}}$, we obtain $n_{s}\simeq 0.07 \times 10^{20}$\,cm$^{-3}$, 
far below the reported $n_{n}$ value. A similar mismatch was observed in TSCs showing a nematic character in the superconducting state, such as Cu$_{x}$Bi$_{2}$Se$_{3}$ and Sr$_{x}$Bi$_{2}$Se$_{3}$ \cite{krieger2018,Amato2018}. For instance, in Sr$_{x}$Bi$_{2}$Se$_{3}$ single crystals, $\lambda_{0}\simeq 2.3$\,$\mu$m \cite{Amato2018}. Hence, by assuming an effective mass of the charge carriers equal to the electron mass, one obtains $n_{s}\simeq 0.05 \times 10^{20}$\,cm$^{-3}$, clearly inconsistent with $n_{n}\sim 1.2 \times 20^{20}$\,cm$^{-3}$, the normal-state carrier density from low-temperature Hall-effect measurements, as reported in Ref.~\cite{Amato2018}. We believe that such inconsistency between $n_{s}$ and $n_{n}$ may be a common feature shared by all nematic superconductors.

Finally, we briefly comment on the possible superconducting order parameter 
realized in {\ca}. By using group-theory classifications, we can determine 
possible pairing symmetries, which induce a spontaneous rotational-symmetry 
breaking in the superconducting state of {\ca}. By discarding unlikely 
pairing states higher than the $f$-wave, we consider the irreducible representations 
of the {\oh} group. Then, as shown in Table \ref{tab:GT} (see also Ref.~\onlinecite{sigri91}),  
we classify the possible superconducting pairing states with a lower 
symmetry, here with a point group $D_{4h}$ (a subgroup of $O_{h}$),
which break the rotational crystallographic symmetry.
Here, we assume that no admixture of pairing states is realized in {\ca}. 
Although further parity-sensitive experiments, including investigations 
of the NMR Knight shift, are required, the fully-gapped superconducting 
state without time-reversal-symmetry breaking unveiled by our 
{\musr} measurements, together with the nematicity of {\ca}, 
suggests that a \emph{fully-gapped} \textit{d}-vector,  
$\boldsymbol{d}(\boldsymbol{k}) = 2\hat{\boldsymbol{z}}k_{z} - 
\hat{\boldsymbol{x}}k_{x} - \hat{\boldsymbol{y}}k_{y}$, 
in the two-dimensional representation $E_{u}$ can possibly be realized
in {\ca} (see Table ~\ref{tab:GT}). This odd-parity
state with a fully opened gap satisfies one of the key 
prerequisites for topological superconductivity in centrosymmetric 
time-reversal-invariant systems \cite{fu10}. Along with the odd number 
of TRIM enclosed by the Fermi surfaces previously reported~\cite{siddi21}, 
{\ca} turns out to be a promising candidate material to exhibit 
topological superconductivity.

\subsection{Conclusion}
In conclusion, we investigated the magnetotransport and magnetic 
properties of the superconducting topological semimetal {\ca}. 
We observe an unusual twofold symmetry of {\hc2} about a fourfold 
symmetric axis and an anomalous temperature dependence of the upper 
critical field, both suggestive of unconventional pairing. {\musr} 
measurements confirm the breaking of the rotational symmetry also with 
respect to magnetic properties, indicative of the realization of 
nematic superconductivity. 
Since \zfmu\ evidences a preserved time-reversal symmetry, 
while \tfmu\ reveals a fully-gapped superconducting state, 
we can restrict a possible superconducting order parameter 
in {\ca} to the \textit{d}-vector $\boldsymbol{d}(\boldsymbol{k}) = %
2\hat{\boldsymbol{z}}k_{z} - \hat{\boldsymbol{x}}k_{x} - %
\hat{\boldsymbol{y}}k_{y}$.
Overall, our results suggest that the cubic {\ca} system fulfills one of the proposed 
criteria for topological superconductivity. A theoretical study addressing in full details the superconducting state emerging in {\ca} is planned for the next future.

\begin{acknowledgments}
The authors thank K.\ Izawa, R.\ Klemm, and Y.\ Nagai for helpful discussions. This work was supported by startup funding from the University of Central Florida. H.S.\ and Y.N.\ were supported by NSF CAREER DMR-1944975. P.V., C.N., and E.D.B.\ were supported by the US NSF under Grant No.\ DMR-1503627. Part of this work was performed at the Swiss Muon Source S$\mu$S, Paul Scherrer Institut, Villigen, Switzerland. T.S. was supported by the Swiss National Science Foundation (SNSF) (Grant No.\ 200021-169455).
\end{acknowledgments}


\appendix{}

\section{{\musr} experimental set up}\label{app:A}

\begin{figure}[tbh]
	\centering
	\includegraphics[width=0.45\textwidth]{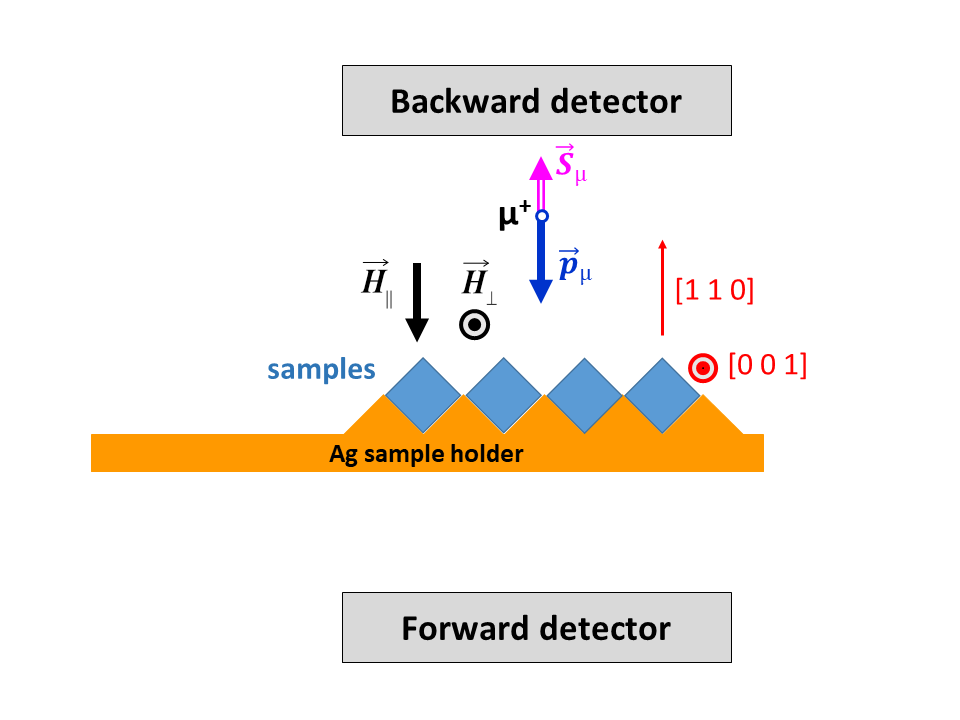} 
	\caption{\label{fig:setup}Outline of the sample holder and relative positions of the positron detectors, muon beam, and applied magnetic
	field (WEU magnet -- $\boldsymbol{H}_{\parallel}$, WEV magnet -- $\boldsymbol{H}_{\bot}$). The red arrow and dot indicate the crystal [110] and [001] directions, respectively.} 
\end{figure}

We realized a sample holder especially conceived for {\ca} single crystals. In Fig.~\ref{fig:setup} we show a transverse section of the sample holder, including the relative directions of the muon beam and the applied magnetic fields. 
The sample holder consists of high-purity (99.99\%) Ag plate, where a series of 
$45^{\circ}$ grooves are etched, covering an area of ca.\ $2 \times 2$\,cm$^2$. 
Since \casn\ has a cubic structure, the groves allowed us to align its  
[110] direction (diagonal of the faces of a cube) either parallel 
or orthogonal to the applied magnetic field.
As shown in Fig.~\ref{fig:setup}, the incident muon momentum 
$\boldsymbol{p}_\mu$ was parallel to the [110] direction. In such configuration, the nematicity of CaSn$_3$ can be
probed simply by changing the orientation of the applied magnetic field
$\boldsymbol{H}$ with respect to the crystal axes.
Two possibilities were used: (a) $\boldsymbol{H}$ parallel   
($H_{\parallel} \neq 0$), or (b) $\boldsymbol{H}$ orthogonal 
($H_{\bot} \neq 0$) to the [110] direction. We define such configurations as ``\emph{parallel}'' and ``\emph{orthogonal}'', 
respectively. The muon spin was rotated only when the applied field was $H_{\parallel}$ (WEU magnets).
The \casn\ single crystals, depicted in blue in 
Fig.~\ref{fig:setup}, were glued by using GE varnish \cite{GE}. 
The resulting mosaic covered an area of about $6 \times 6$\,mm$^2$. 
Due to the high reactivity of \casn\ towards oxygen and moisture, this 
operation was performed in a glove box under controlled atmosphere 
(very pure argon). The sample holder with the oriented crystals 
was then sealed in a glass tube by using a glass plug covered with silicone grease. 
Since the sealing, too, was performed in a glove box, the samples
reached the measuring site (PSI, Switzerland) without ever being exposed
to air. The glass tube was opened to air for only ca.\ 2 minutes in order to
mount the sample holder to the $^{3}$He low-temperature
insert, which was then quickly immersed in pure He gas and cooled down. 
Note that, despite the short exposure to air, the thin GE varnish layer 
covering the sample acts as a very effective protection against 
oxygen and moisture.

\begin{figure*}[tbh]
	\centering
	\includegraphics[width=0.457\textwidth]{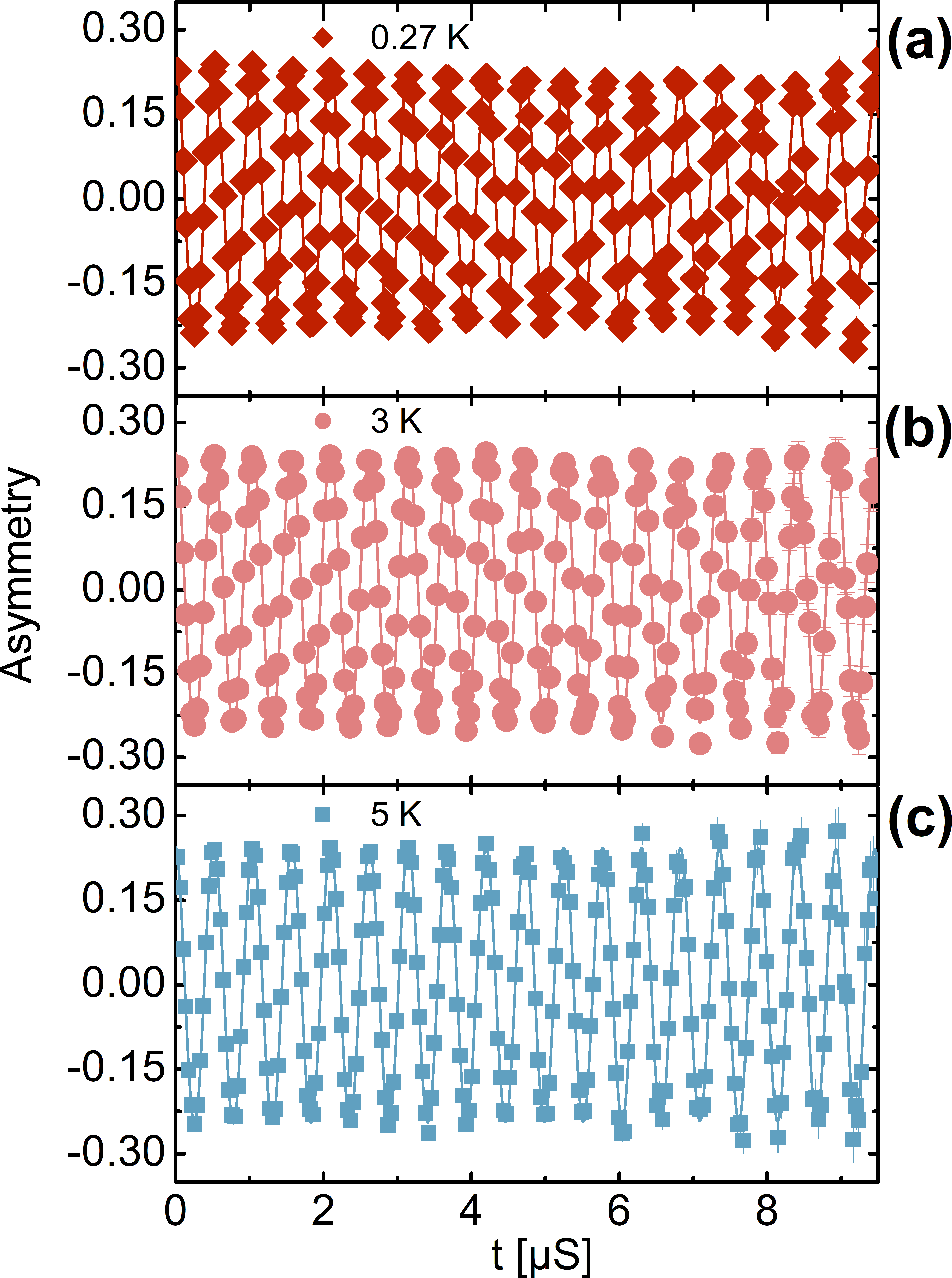} 
	\hspace{5mm}
	\includegraphics[width=0.40\textwidth]{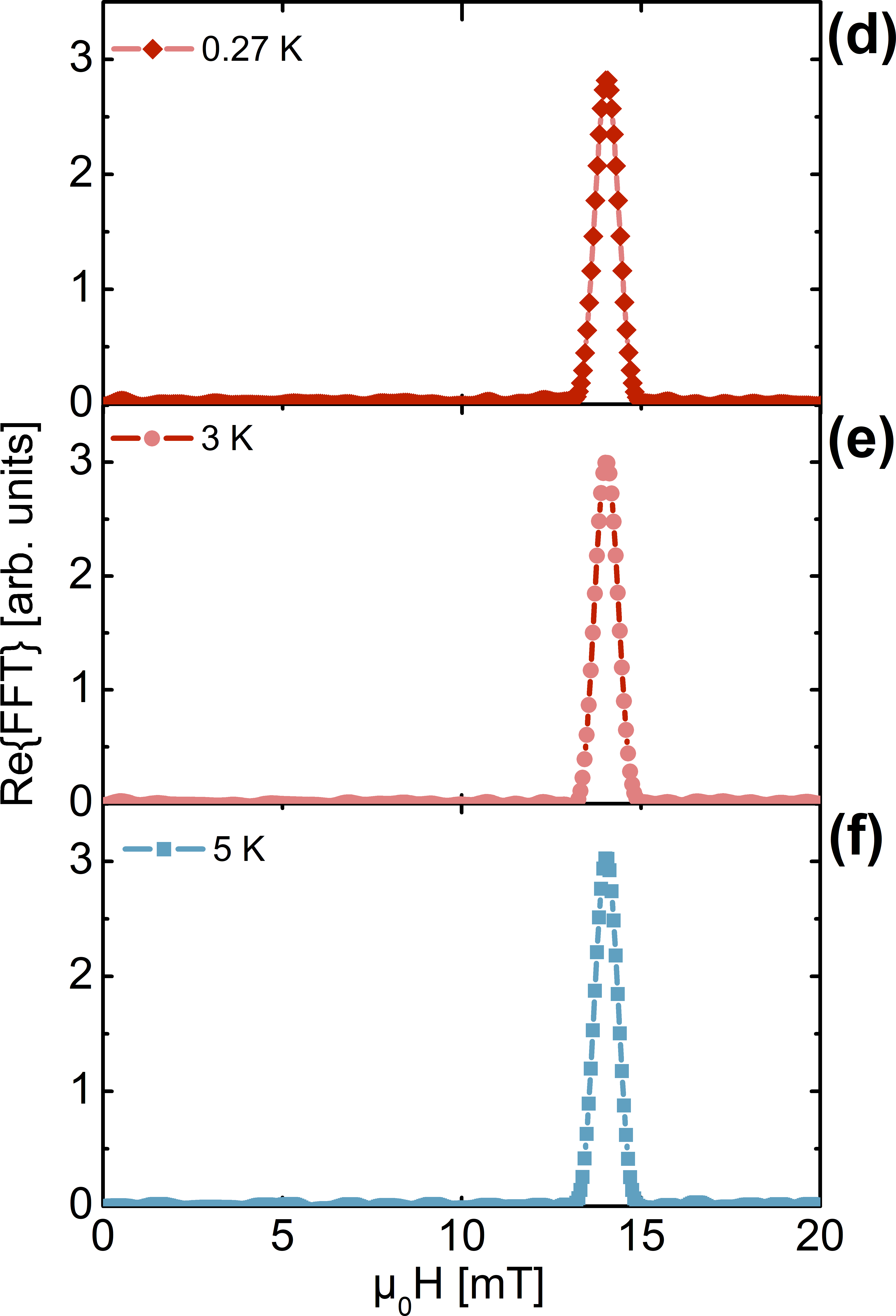} 
	\caption{\label{fig:musrtest}(a,b,c) $\mu$SR asymmetry measured at 0.27\,K (FC), 3\,K (FC), and 5 K, respectively, 
	in an applied field $\mu_{0}H_{\perp}= 14$\,mT. 
	(d,e,f) The corresponding amplitudes of the real part of FFT.} 
\end{figure*}

\begin{figure*}[bht]
\includegraphics[width=0.9\textwidth]{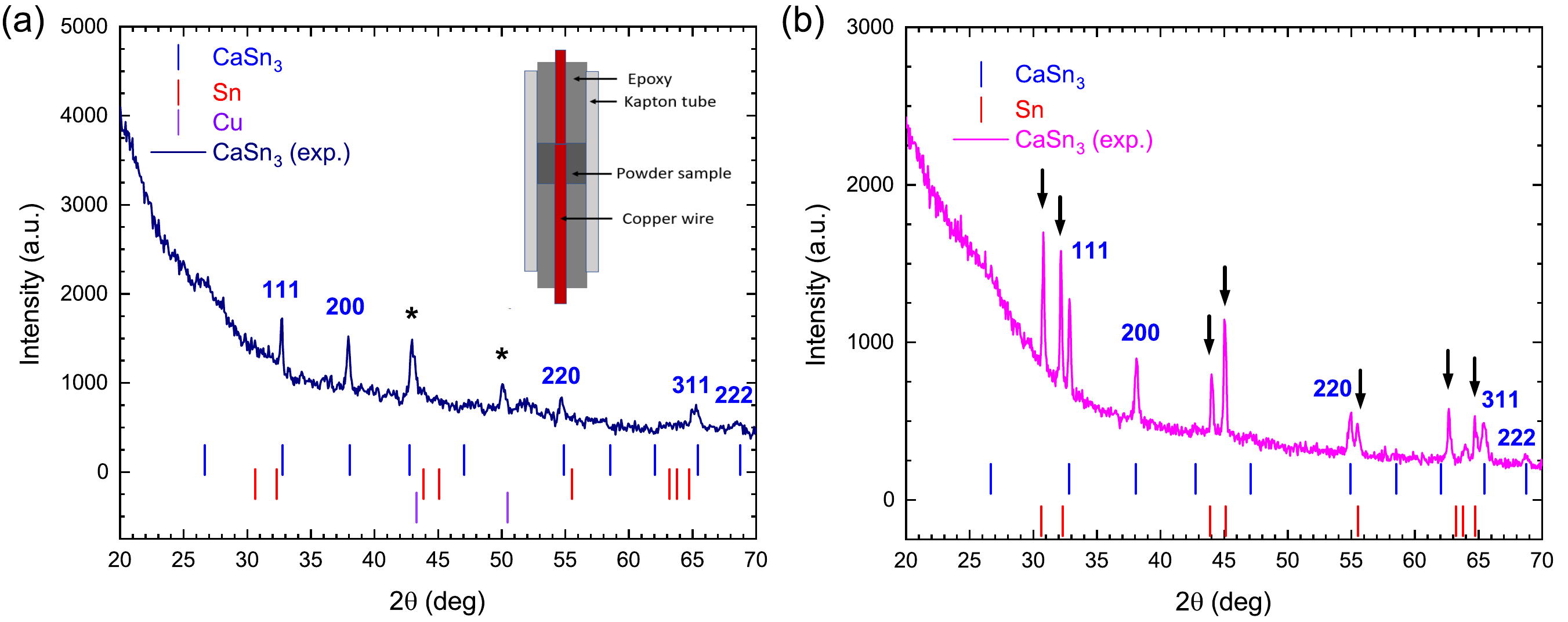}
\caption{\label{fig:xrd}X-ray diffraction pattern of {\ca} obtained using Cu-$K_{\alpha}$ radiation. (a) A powder sample sealed in a Kapton tube with Devcon rapid epoxy.  Asterisks indicate peaks from the Cu wires (here acting as a thermal link). No Sn peaks are observed. (b) A powder sample covered with a Kapton tape. Arrows indicate peaks from decomposed tin. Similar to previous 
reports \cite{luo15a,zhu19}, clear degradation is observed, 
indicating that {\ca} is highly air-sensitive.}
\end{figure*}

\section{{\musr} test on sample degradation}\label{app:B}

\casn\ is particularly sensitive to oxygen and moisture, which can induce the formation of nanoscopic tin inclusions. 
The effect is so strong that, after some hours of exposure to air, the x-ray diffraction pattern shows only Sn peaks. Hence, to ensure 
that \musr\ data refer to a pristine sample, it is mandatory to keep its possible degradation under strict control.

Should the sample degrade during its insertion 
in the Heliox cryostat for the \musr\ measurements, two outcomes are 
possible. The sample may end up as a homogeneous mixture of 
(i) pure-, or (ii) dirty tin (i.e., tin with diluted calcium inclusions).\\
(i) By recalling that pure tin is a type-I superconductor, experiments in 
an applied field $\mu_{0}H_{\perp}= 14$\,mT would show 
peculiar features, depending if tin is in its pure diamagnetic, 
or in the intermediate state \cite{Rustem2019}. As to the former, the local field at the muon implantation sites should be nearly zero with a consequent absence of oscillations in the time dependent asymmetry and the presence of a zero-field peak in the corresponding FFT spectrum. As to the latter, a macroscopic separation into normal- and superconducting mesoscopic domains occurs. 
Thus, the time dependent polarization would consist of an oscillating term 
(muons stopped in the normal regions) superposed to a constant term (muons stopped in the Meissner regions). In this case, FFT spectra would show two peaks, one at zero field (the superconducting fraction) and one at the thermodynamic 
critical field corresponding to the current temperature (the normal domains). However, as shown in 
Fig.~\ref{fig:musrtest}, this is not our case.\\ 
(ii) It is well known that a type-I superconductor turns into a dirty type-II one, upon chemical substitutions and/or inclusions \cite{dirtySn}. As deduced from x-ray measurements on degraded {\ca}, one can classify such system as a tin matrix with diluted Ca content. Such alloy should behave as dirty type-II superconductor, which \emph{preserves} the isotropic character of pure tin. This is clearly not our case, since we record different types of SC behavior depending of the direction of the applied magnetic field.\\
Therefore, in view of the above arguments, the {\ca} sample 
studied here via \musr\ can be considered pure.

\section{X-ray diffraction}\label{app:C}

In Fig.~\ref{fig:xrd}a we show the powder x-ray diffraction pattern obtained from the sample prepared for the magnetic 
ac-susceptibility measurements. The sample preparation procedure is explained in detail in Sec.~\ref{ssec:magn_susc}, while  
the sample holder used for the ac measurements is shown in the inset of Fig.~\ref{fig:xrd}a. No discernible peaks from remaining 
Sn flux or from decomposed Sn are observed. For a comparison, in Fig.~\ref{fig:xrd}b we show the powder 
x-ray diffraction pattern obtained on a sample prepared following the procedure described in Ref.~\onlinecite{zhu19}. In this case, we note the presence of Sn peaks, a clear signature of sample degradation.

\bibliographystyle{apsrev4-2}

\begin{thebibliography}{55}
\bibitem [{Nfa()}]{Nfactor}%
  \BibitemOpen
  \href@noop {} {}\bibinfo {note} {The magnetic susceptibility was calculated
  by assuming a demagnetizing factor $N=1/3$, i.e., by considering the powder
  sample to consists of spherical grains.}\BibitemShut {Stop}%
\bibitem [{GE()}]{GE}%
  \BibitemOpen
  \href@noop {} {}\bibinfo {note} {{B}efore using the {GE} varnish, we checked
  that it does not degrade the as-grown {CaSn}$_3$ single
  crystals.}\BibitemShut {Stop}%
\end{thebibliography}

\end{document}